\documentclass[conference]{IEEEtran}

\IEEEoverridecommandlockouts
\IEEEaftertitletext{\vspace{-2\baselineskip}}
\usepackage{amsmath,graphicx}
\usepackage{acronym}
\usepackage{url}
\usepackage{subfig,balance}
\usepackage{tabularx}
\usepackage{booktabs,multirow} 
\acrodef{TF}{Time-Frequency}
\acrodef{HTS-AT}{Hierarchical Token-semantic Audio Transformer}
\usepackage{booktabs}
\usepackage{amssymb}
\usepackage{balance}
\usepackage{caption}

\acrodef{RIR}{Room Impulse Response}
\acrodef{ER}{Emotion Recognition}
\acrodef{MER}{Multi-modal Emotion Recognition}
\acrodef{HRI}{Human-Robot Interaction}
\acrodef{CREMA-D}{Crowd-sourced Emotional Multimodal Actors Dataset}

\acrodef{NLP}{Natural Language Processing}
\acrodef{AST}{Audio Spectrogram Transformer}
\acrodef{LSTM}{Long Short-Term Memory}
\acrodef{BLSTM}{Bidirectional Long Short-Term Memory}
\acrodef{RNN}{Recurrent Neural Networks}
\acrodef{CNN}{Convolutional Neural Networks}

\acrodef{SNR}{signal-to-noise ratio}
\acrodef{ViT}{Vision Transformer}
\acrodef{CLS}{Class Token}
\acrodef{MHA}{multi-head attention}
\acrodef{TCN}{Temporal Convolutional Networks}

\acrodef{ACE}{Acoustic Characterisation of Environments}
\acrodef{CE}{Cross-Entropy}
\acrodef{SOTA}{state-of-the-art}
\acrodef{AST}{Audio Spectrogram Transformer}
\acrodef{ASR}{Automatic Speech Recognition}
\acrodef{BESR}{Binaural Emotional Speech Recognition}
\acrodef{HRI}{Human-Robot Interaction}
\acrodef{SER}{Speech Emotion Recognition}
\acrodef{CNN}{Convolutional Neural Networks}
\acrodef{DNN}{Deep Neural Networks}
\acrodef{RNN}{Recurrent Neural Networks}
\acrodef{BiLSTM}{Bidirectional Long Short-term Memory}
\acrodef{LSTM}{Long Short-term Memory}
\acrodef{MFCC}{mel frequency cepstral coefficients}
\acrodef{FC}{fully-connected}
\acrodef{FFT}{fast Fourier transform}
\acrodef{STFT}{short-time Fourier transform}
\acrodef{t-SNE}{t-distributed Stochastic Neighbor Embedding}
\acrodef{RAVDESS}{Ryerson audio-visual database of emotional speech and song}
\acrodef{IEMOCAP}{Interactive Emotional Dyadic Motion Capture}

\acrodef{GloVe}{Global Vectors for Word Representation}

\title{Multi-Microphone and Multi-Modal Emotion Recognition in Reverberant Environment\thanks{The project has received funding from the European Union’s Horizon 2020 Research and Innovation Programme, Grant Agreement No. 871245; and  from the Audition Project, Data Science Program, Council of Higher Education, Israel.}
}
\author{\IEEEauthorblockN{Ohad Cohen}
\IEEEauthorblockA{\textit{Faculty of Engineering} \\
\textit{Bar-Ilan University}\\
Ramat-Gan, Israel \\
ohad.cohen@biu.ac.il\\
0009-0002-4707-881X}
\and
\IEEEauthorblockN{Gershon Hazan}
\IEEEauthorblockA{\textit{Faculty of Engineering} \\
\textit{Bar-Ilan University}\\
Ramat-Gan, Israel \\
gershon.hazan@biu.ac.il}
\and
\IEEEauthorblockN{Sharon Gannot}
\IEEEauthorblockA{\textit{Faculty of Engineering} \\
\textit{Bar-Ilan University}\\
Ramat-Gan, Israel \\
sharon.gannot@biu.ac.il\\ 0000-0002-2885-170X}
}

\def\BibTeX{{\rm B\kern-.05em{\sc i\kern-.025em b}\kern-.08em
    T\kern-.1667em\lower.7ex\hbox{E}\kern-.125emX}}
\begin{document}
\bstctlcite{IEEEexample:BSTcontrol}
%
\maketitle
\begin{abstract}
This paper presents a \ac{MER} system designed to enhance emotion recognition accuracy in challenging acoustic conditions. Our approach combines a modified and extended \ac{HTS-AT} for multi-channel audio processing with an $R(2+1)$D \ac{CNN} model for video analysis. We trained and evaluated our proposed method on a reverberated version of the \ac{RAVDESS} dataset using synthetic and real-world \acp{RIR}. 
Our results demonstrate that integrating audio and video modalities yields superior performance compared to uni-modal approaches, especially in challenging acoustic conditions.
Moreover, we show that the multimodal (audiovisual) approach that utilizes multiple microphones outperforms its single-microphone counterpart. 
\end{abstract}
\begin{IEEEkeywords}
Emotion recognition, multi-modal, reverberant conditions, audio transformer
\end{IEEEkeywords}
\section{Introduction}
\label{sec:intro}

\ac{ER} is a crucial component in human-computer interaction, with applications ranging from healthcare to customer service. Humans naturally express emotions across multiple modalities, including facial expressions, language, speech, and gestures. Accurately modeling the interactions between these modalities, which contain complementary and potentially redundant information, is essential for effective emotion recognition. Most existing studies primarily focus on uni-modal emotion recognition, concentrating on either text, speech, or video \cite{bharti2022text,2022dalia,zhou2023emotion}. Although significant advancements in single-modal emotion recognition have been demonstrated, these models often fall short in complex scenarios since they do not utilize the inherently multi-modal nature of emotional expression. Moreover, research on jointly employing multi-modal and multi-microphones for \ac{ER} is relatively scarce.
Previous works have made significant strides in \ac{MER}. Studies such as \cite{john2022audio,noroozi2017audio, dai2021multimodal, kansizoglou2019active} have developed systems that simultaneously analyze visual and acoustic data. In \cite{franceschini2022multimodal}, researchers presented an unsupervised \ac{MER} feature learning approach incorporating audio-visual and textual information. 
These studies often overlooked the challenges posed by real-world acoustic conditions, particularly reverberation and noise, which can significantly impact the performance of audio-based emotion recognition. 
Feature selection is vital in designing effective \ac{MER} systems. For acoustic features, log-mel filterbank energies and log-mel spectrograms have been widely adopted \cite{huang2017deep,seo2020fusing}. In the video domain, various deep learning architectures such as VGG16 \cite{song2021video}, I3D \cite{ghaleb2019multimodal}, and FaceNet \cite{schroff2015facenet} have been employed, along with facial features like landmarks and action units extracted using tools like OpenFace \cite{zadeh2018multimodal}. For the text modality, \ac{GloVe} \cite{pennington2014glove} have been frequently used \cite{zadeh2018multimodal, mittal2020m3er,delbrouck2020modulated}. 

Despite these advancements, a gap remains in addressing the challenges posed by reverberant and noisy environments. Real-world acoustic conditions can significantly alter speech signals, potentially degrading the performance of audio-based emotion recognition systems. Moreover, the integration of multi-channel audio processing techniques with video analysis for emotion recognition in such challenging conditions has not been thoroughly explored. 

This work addresses these limitations by proposing a \ac{MER} with multi-channel audio that outperforms solutions solely based on single-channel audio. We propose a novel approach that combines two state-of-the-art architectures for audio-visual emotion recognition. The multi-channel extension of the  \ac{HTS-AT} architecture for the audio modality \cite{ohad2024} and the $R(2+1)$D \ac{CNN} model \cite{tran2018closer} for the video modality. 
We use a reverberated version of the \ac{RAVDESS} \cite{livingstone2018ryerson} dataset to analyze the proposed scheme's performance.  Reverberation was added by convolving the speech utterances with real-life \acp{RIR} drawn from the \ac{ACE} challenge dataset \cite{7486010}. The code of the proposed method is available.\footnote{\texttt{ https://github.com/OhadCohen97/Multi-Microphone-\\Multi-Modal-Emotion-Recognition-in-Reverberant-\\Environments.}}

\section{Problem Formulation}
\label{sec:Problem Formulation}
Denote the two modalities as $ M = \{\text{video}, \text{audio}\}$ and the set of emotions as: 
\begin{multline}
    E = \{\text{happy},\text{calm}, \text{sad}, \text{angry},\ldots\\ \text{neutral}, \text{fearful},\text{disgust},\text{surprised} \}.
\end{multline}
Let $v(t)$ be the video signal and $s(t)$ the anechoic audio signal, with $t$ the time index. An array of $C$ microphones captures the audio signal after propagating in the acoustic environment. The signals, as captured by the microphones, are given by:
\begin{equation}
    y_i(t) = \{s * h_i\}(t), \; i = 1,2,\ldots,C,
\end{equation}
 where $h_i(t),\; i = 1,2,\ldots,C,$ are the \acp{RIR} from the source to the $i$th microphone. The feature embeddings for each modality are denoted  $f_v$ and $f_s$, respectively. This study aims to classify the utterance to one of the emotions using the available information and utilizing the relations between the feature embeddings of both modalities:
 \begin{equation}
     M\left\{v(t), \{y_i(t)\}_{i=1}^C\right\} \Rightarrow f_v \oplus f_s \Rightarrow E, 
 \end{equation}
where $\oplus$ stands for late fusion concatenation.
%

\begin{figure*}[t]
    \centering
      \includegraphics[width=0.85\textwidth]{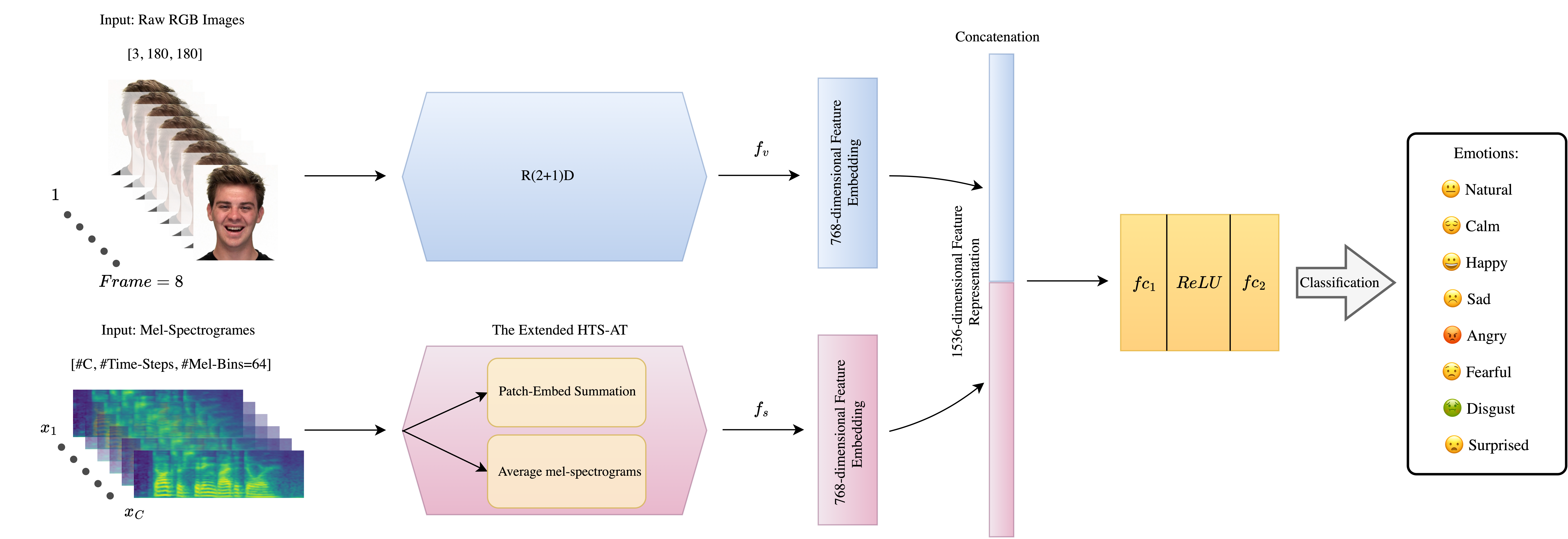}
      \caption{The proposed \acf{MER}.}
      \label{Mer}
   \end{figure*}
\section{Proposed Model}
\label{section:Proposed_Model}
Our proposed \ac{MER} architecture combines two powerful models: the modified and extended \ac{HTS-AT} \cite{ohad2024} for multi-channel audio processing and the $R(2+1)$D model \cite{tran2018closer} for video analysis. These uni-modal models are integrated to create a robust multi-modal system for emotion recognition in challenging acoustic conditions.

The input features of the models are the mel-spectrograms for the audio track and the raw RGB facial images for the visual track. For the audio modality, we followed the same preprocessing procedure as in \cite{ohad2024} and used the SpecAugment Library \cite{park2019specaugment} to augment the mel-spectrograms. For augmenting the video modality, we used the TorchVision Library \cite{torchvision2016}.

\noindent\textbf{Audio:} The extended multi-channel \ac{HTS-AT} model addresses the integration of multi-microphone information, employing the Swin-Transformer architecture \cite{liu2021swin},\footnote{\texttt{github.com/microsoft/Swin-Transformer}} a variant of the \ac{ViT} \cite{dosovitskiy2020image} architecture. The architecture is also applicable to the single-microphone configurations, namely $C=1$. The model's architecture consists of four groups, each comprising Swin-Transformer blocks with varying depths. In addition, the model uses a hierarchical structure and windowed attention mechanism to efficiently process mel-spectrograms, which serve as our audio feature extractor. We use the two multi-channel variants with the modified \ac{HTS-AT} module, as proposed in \cite{ohad2024}: 1) Patch-Embed Summation - the mel-spectrogram of each channel is processed through a shared Patch-Embed layer, after which the outputs are summed across channels; and 2) Average mel-spectrograms - mel-spectrograms from multiple channels are averaged before being fed into the model. More details can be found in \cite{ohad2024}.

\noindent\textbf{Video:} For video feature extraction, we employ the pre-trained $R(2+1)$D model, an 18-layer ResNet-based architecture designed for action recognition. The $R(2+1)$D model decomposes the 3D convolutions into separate spatial (2D) and temporal (1D) convolutions, which allows it to effectively capture both spatial and temporal features in the video data. Fig.~\ref{Mer} presents the integration of the two modalities.

\noindent\textbf{Feature Concatenation:} The feature embeddings are extracted from the extended multi-channel HTS-AT and the $R(2+1)$D models, followed by concatenation to create a unified multi-modal representation. This combined feature vector captures audio and visual cues relevant to emotion recognition. The concatenated features are then passed through two fully connected layers for final classification. These layers learn to interpret the combined audio-visual features and to map them to emotion categories. The output of the final layer corresponds to the predicted emotion class. 
This integrated scheme ensures that the multi-channel audio and visual data are effectively processed and leveraged. This allows the model to capture and utilize complementary information from both modalities, thus achieving improved \ac{ER} accuracy.

\section{Experimental Study}
\label{Experimental study}
This section outlines the experimental setup and describes the comparative analysis between the proposed scheme and a baseline method.
\subsection{Datasets} Our work utilized the \ac{RAVDESS} dataset for emotion recognition. This dataset includes 24 actors, equally divided between male and female speakers, each delivering 60 English sentences. Hence, there are 1440 audio-video pairs representing eight different emotions (`sad,' `happy,' `angry,' `calm,' `fearful,' `surprised,' `neutral,' and `disgust'). All utterances are pre-transcribed. Therefore, the emotions are expressed more artificially compared to spontaneous conversation. The RAVDESS dataset is balanced across most classes except for the neutral class, which has a relatively small number of utterances. We used an \textit{actor-split} approach, dividing the data into 80\% training, 10\% validation, and 10\% test sets, ensuring no actor appears in more than one split. As a result, model accuracy may be lower than reported in some prior works because the test set includes actors not seen during fine-tuning.
As publicly available multi-microphone datasets for \ac{SER} do not exist, we generated our own
dataset.
We used synthesized \acp{RIR} to fine-tune the multi-channel experiment model. We employed the `gpuRIR` Python package\footnote{\texttt{github.com/DavidDiazGuerra/gpuRIR}} to simulate reverberant multi-channel microphone signals (setting the number of microphones to $C=3$). Each clean audio sample from the \ac{RAVDESS} dataset was convolved with distinct multi-channel \acp{RIR}, resulting in 1440 3-microphone audio samples. The associated video data is unaffected by reverberation. 
We simulated rooms with lengths and widths uniformly distributed between 3~m and 8~m, maintaining a constant height of 2.9~m and an aspect ratio between 1 and 1.6. We randomly positioned the sound source and microphones within these simulated environments under the following constraints. The sound source was placed at a fixed height of 1.75~m, with its $x$ and $y$ coordinates randomly determined within the room, ensuring a minimum distance of 0.5~m from the room walls. Similarly, the microphones were positioned at a fixed height of 1.6~m, with their $x$ and $y$ coordinates also randomly determined within the room dimensions. The reverberation time was set at the range $T_{60}=500-850$~ms. The distance between the sound source and microphones was randomly selected in the range $[0.2,d_c]$~m, where $d_c$ to the critical distance, determined by the room’s volume and $T_{60}$. This ensures the dominance of direct sound over reflections.
Finally, we added spatially-white noise with \ac{SNR} of $20$~dB to each reverberant signal. This noise was synthesized by applying an auto-regressive filter of order 1 to a white Gaussian noise, emphasizing lower frequencies. 

The proposed scheme was evaluated using real-world \acp{RIR} drawn from the \ac{ACE}  database \cite{7486010}. The \ac{ACE} \ac{RIR} database comprises recordings from seven different rooms with varying dimensions and reverberation levels (see Table~\ref{ACE_results}). We only used a subset of the database, recorded with a mobile phone equipped with three microphones in the near-field scenario, which is a practical choice for real-world \ac{SER} applications. We convolved all audio utterances of the test set with the \ac{ACE} \acp{RIR} to generate 3-microphone signals for each utterance.

\subsection{Algorithm Setup}
\label{Setup}
\sloppy
As discussed earlier, the video modality leverages the $R(2+1)$D model pre-trained on the action recognition Kinetics dataset \cite{kay2017kinetics}. To better suit our \ac{ER} task, we modified the model's architecture by adjusting the final linear layer. Specifically, we reconfigured it to output 768-dimensional feature embeddings. This adjustment ensures that both modalities (video and audio) contribute equally-sized feature vectors to the multi-modal representation by fusion through concatenation. 

The resolution of the RGB video frames was first reduced to $224 \times 224$ pixels. Then, eight frames from the video stream were randomly selected. To augment the dataset, these frames underwent refinement through random cropping using the TorchVision Library \cite{torchvision2016}, yielding $180 \times 180$ images that enhance the model's robustness to spatial variations. In addition, we added random horizontal and vertical flips, each with a 30\% probability of application, coupled with arbitrary rotations within the range of [-30$^\circ$, 30$^\circ$]. 

The audio modality applies an extended version of the \ac{HTS-AT} model \cite{ohad2024}, suitable for both multi-channel and single-channel scenarios. As described in Sec.~\ref{section:Proposed_Model}, the network structure configuration is arranged into four groups, each containing several Swin-Transformer blocks: 2, 2, 6, and 2, respectively. The mel-spectrogram input is initially transformed into patches and linearly projected to a dimension of $D = 96$. This dimension expands exponentially through each transformer group, finally reaching a dimension of $768$ ($8D = 768$), which matches the design principles of \ac{AST}. Pre-processing was carried out as explained in \cite{ohad2024} both for multi-channel and single-channel experiments. We augmented the mel-spectrograms by using the SpecAugment Library \cite{park2019specaugment}, which consists of temporal masking, occluding four distinct ``strips'', each 64 time-frames long. Complementing this, we applied frequency domain masking, obscuring two strips, each 8 frequency bands wide.

Our multi-modal approach combines the feature embeddings from both the video and audio modalities. The 768-dimensional feature vectors extracted from the $R(2+1)$D model and the extended \ac{HTS-AT} model are concatenated, resulting in a 1536-dimensional feature representation. This combined feature vector is then fed into a classification head for prediction. The right-hand side of Fig.~\ref{Mer} presents two fully connected layers $(fc)$ with a Relu activation function between them, forming the sequence:
\begin{equation}
     f_{c_1} \to \text{ReLU} \to f_{c_2}\ \Rightarrow E
\end{equation}

The fine-tuning processes are applied using the Adam optimizer with a learning rate of $1e^{-3}$ and a warm-up strategy. We used cross-entropy loss as the metric with a batch size of 32. The maximum number of epochs was set to 500 for all experiments, with an early stopping strategy with a patience of 12 to prevent overfitting. In practice, the maximum number of epochs was never reached, as the fine-tuning process was halted earlier due to the activation of the patience parameter. The overall number of parameters for the fine-tuned models are as follows: 32.3M for the uni-modal scheme based on video, 28.7M for the uni-modal scheme based on audio, and 62.7M for the multi-modal scheme.
\subsection{Results}
\label{Results}
\begin{table}[htbp]
    \centering
        \caption{The Accuracy results of single-microphone \ac{MER} method compared with SOTA MER methods tested on the original RAVDESS dataset. Results for the competing methods are taken from the corresponding articles.}
    \label{results_sota}
    \renewcommand{\arraystretch}{1.2}
    \begin{tabular}{l c}
        \toprule
        \textbf{Methods} & \textbf{ACC (\%)} \\
        \midrule
        Human performance \cite{livingstone2018ryerson} & 80.00 \\
        \midrule
        Garaiman et al. \cite{garaiman2024multimodal} & 65.76 \\
        Ghaleb et al. \cite{ghaleb2020multimodal} & 76.30 \\
        Franceschini et al. \cite{franceschini2022multimodal} & 78.54 \\
        Radoi et al. \cite{radoi2021end} & 78.70 \\
        \textbf{Luna-Jim{\'e}nez et al.} \cite{luna2021multimodal} & \textbf{80.08} \\
        Proposed \ac{MER} ($C=1$) & 80.00 \\
        \bottomrule
    \end{tabular}
    \vspace{-0.6cm}
\end{table}
Table~\ref{results_sota} compares the performance of the proposed \ac{MER} approach (single-microphone variant, $C=1$) with several \ac{SOTA} \ac{MER} approaches evaluated on \ac{RAVDESS}. The results indicate that our single-microphone \ac{MER} achieves performance on par with \cite{franceschini2022multimodal, radoi2021end, luna2021multimodal}. Moreover, to assess and visualize the separation capabilities of the proposed scheme across the clean \ac{RAVDESS} dataset, we employed the \ac{t-SNE} visualization method. This nonlinear technique reduces high-dimensional data into two- or three-dimensional representations suitable for graphical visualization. Importantly, it maps nearby points in the high-dimensional space to close points in the reduced space, while far-apart points remain distant in the visualization \cite{van2008visualizing}. 
\vspace{-.5cm}
\begin{figure}[htbp]
    \centering
    \qquad\subfloat[\centering RAVDESS: Features]{{\includegraphics[width=0.47\columnwidth]{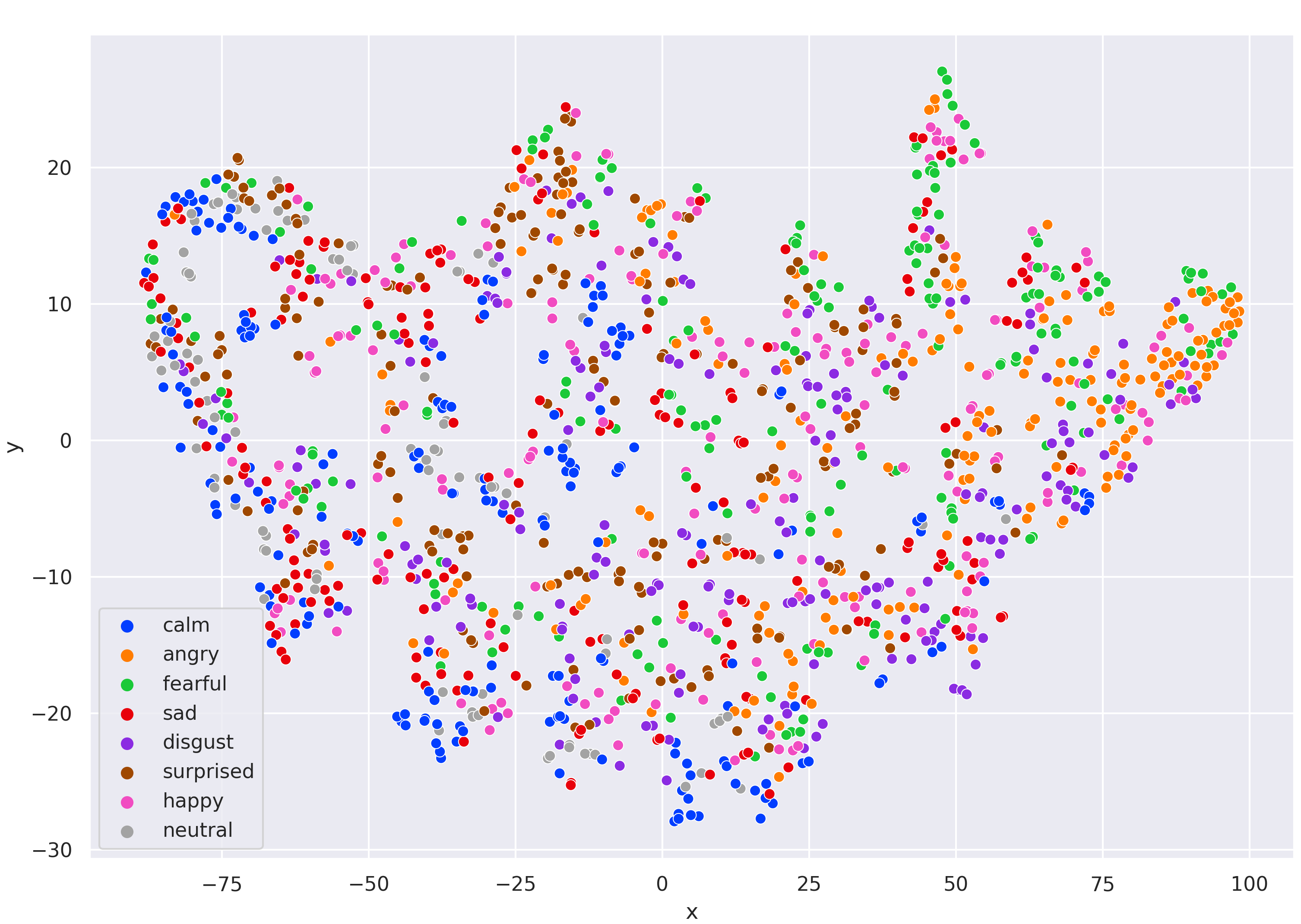} }}%
    \subfloat[\centering RAVDESS: Network output]{{\includegraphics[width=0.47\columnwidth]{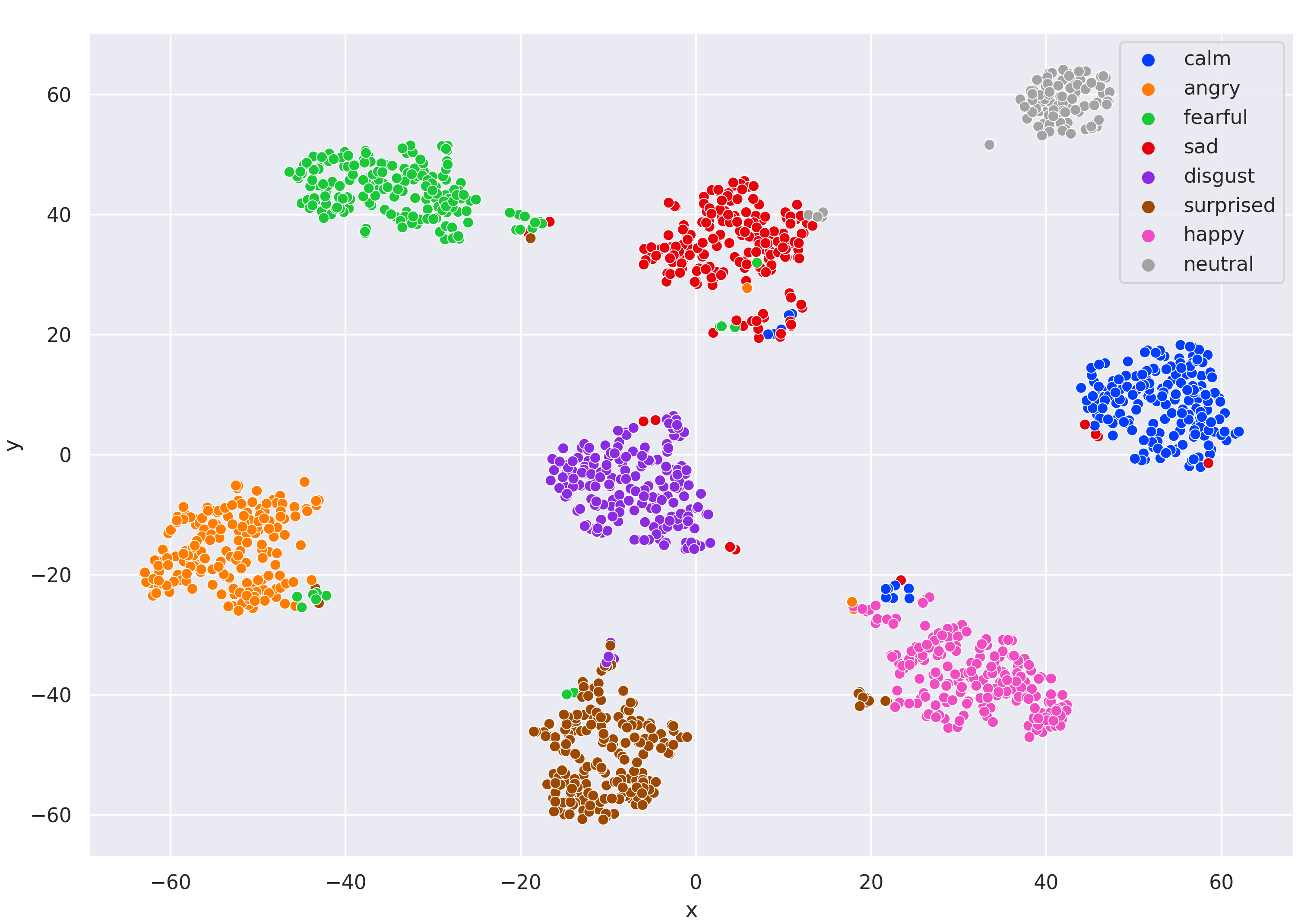} }}%
    \caption{t-SNE visualization.}
    \label{TSNE}  
\end{figure}
In Fig.~\ref{TSNE}, we compare the \ac{t-SNE} mapping of the input features extracted from the \ac{RAVDESS} dataset and the network output, depicting each emotion with a unique color and shape to visualize the clustering quality. The enhancement in classification performance following the network's application is immediately apparent.

We now turn to the evaluation of the multi-channel schemes, using the reverberant \ac{RAVDESS} dataset version, applying \ac{MER} with a multi-microphone ($C=3$). Table~\ref{ACE_results} details our Accuracy results of the various emotion recognition schemes for seven different rooms from the \ac{ACE} database. The video-only modality is compared with the audio-only modality (both single- and multi-channel models) and the combined multi-modal approach. As the video modality is unaffected by the acoustic conditions, we only report the results once. We investigated two single-channel ($C=1$) variants: one fine-tuned on clean speech and the other on reverberant speech. Both were evaluated using a single microphone from the \ac{ACE} test set.
To assess the reliability of our results, we report the mean results together with 75\% confidence intervals.\footnote{\texttt{github.com/luferrer/ConfidenceIntervals}} 
\begin{table*}[t]
  \renewcommand*{\arraystretch}{1.1}
  \centering
    \caption{Accuracy and the associated confidence intervals of the proposed method for the \ac{RAVDESS} test set reverberated by \acp{RIR} drawn from the \ac{ACE} database (using the 3-microphone of the near-filed cellular phone). The `Single-Channel' columns use an arbitrarily chosen microphone, fine-tuned on either clean or reverberant data, respectively. The `Avg mel' columns present results with mel-spectrograms averaged across three channels during fine-tuning and testing. The `Sum PE' columns depict the Patch-Embed fusion approach fine-tuned and tested on the three channels. The asterisk in the video column describes the same result. The best results for each modality are underlined, while the overall best result is shown in boldface.}
    \resizebox{2.03\columnwidth}{!}{
    \begin{tabular}{l ccc cccc cccc}
     \toprule
      \multirow{2}{*}{Room ($T_{60}$~[ms])} &  \multicolumn{3}{c}{Video} & \multicolumn{4}{c}{Audio} & \multicolumn{4}{c}{MER} \\
      \cmidrule(lr){2-4}\cmidrule(r){5-8}\cmidrule(r){9-12}
                           & &                 &     & Clean Single-Channel & Rev Single-Channel &  Avg mel      & Sum PE         & Clean Single-Channel & Rev Single-Channel  & Avg mel     & Sum PE \\
      \midrule
      Lecture  1 ($638$)  & & \underline{69.4} (65.5-73.3)& & 42.7 (38.8-47.2) & 60 (55.5-64.4)   & \underline{61.6} (57.7-65.5) & 61.1 (57.2-65)   & 70 (66.1-73.8)   & 75.0 (71.1-78.8)   & 77.2 (73.3-81.1) & \underline{\textbf{78.3}} (74.4-81.6)  \\
      Lecture  2  ($1220$) & & * &                     & 40.5 (36.6-45)   & 55 (50.5-59.4)   & \underline{60.5} (56.1-64.4) & 58.8 (54.9-62.8) & 71.6 (67.7-75.5) & 76.1 (72.7-79.4) & 76.1 (72.7-80)   & \underline{\textbf{78.3}} (75.0-81.6) \\
      Lobby  ($646$)       & & * &                     & 40.5 (36.6-45)   & 57.2 (53.3-61.1) & 63.3 (59.4-67.22)& \underline{69.4} (65.5-73.3) & 68.3 (64.4-72.2) & 73.3 (69.4-77.2) & 77.2 (73.3-81.1) & \underline{\textbf{77.7}} (74.4-81.1) \\
      Meeting  1 ($437$)   & & * &                     & 42.7 (38.8-46.6) & 57.2 (52.7-61.6) & 60 (56.1-63.8)   & \underline{61.1} (57.2-65)   & 70 (66.1-73.8) & 72.7 (68.8-76.6)   & 77.2 (73.8-81.1) & \underline{\textbf{78.8}} (75.5-82.2) \\
      Meeting  2 ($371$)   & & * &                     & 38.8 (34.4-42.7) & 56.1 (51.6-60.5) & \underline{62.7} (58.8-66.6) & 59.4 (55.5-63.8) & 71.1 (67.2-75) & 75 (71.1-78.8)     & \underline{\textbf{78.8}} (75.5-82.2) & 75.5 (71.6-78.8) \\
      Office 1 ($332$)     & & * &                     & 42.7 (38.3-46.6) & 59.4 (55-63.3)   & 62.7 (58.8-68.3) & \underline{63.3} (58.8-67.2) & 70 (66.1-73.8) & 75 (71.1-78.8)     & \underline{\textbf{77.7}} (74.4-81.6) & 77.2 (73.8-80.5) \\
      Office 2 ($390$)     & & * &                     & 47.7 (43.3-52.2) & 56.6 (52.2-60.5) & \underline{64.4} (60.5-68.3) & 60.5 (56.6-64.4) & 68.8 (65-72.7) & 70 (66.1-73.8)     & 74.4 (70.5-78.3) & \underline{\textbf{75.5}} (71.6-78.8) \\
      \bottomrule
      \label{ACE_results}
    \end{tabular}
    }
    \vspace{-.6cm}
\end{table*}
\begin{figure}[htbp]
    \centering
    \qquad\subfloat[\centering RAVDESS in ACE Office~2 ($T_{60}=390$~ms).]{{\includegraphics[width=0.48\columnwidth]{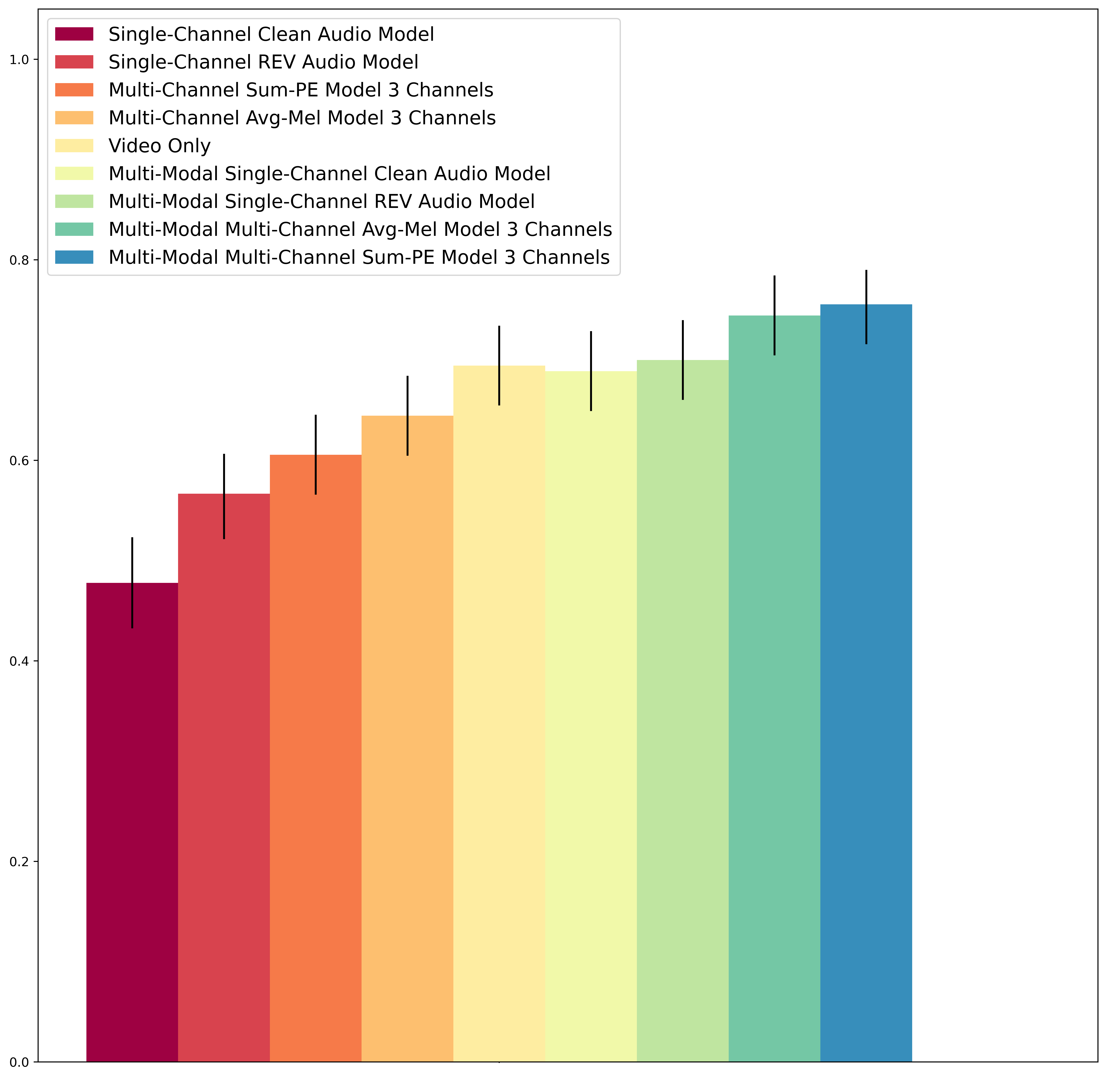} }}%
    \subfloat[\centering RAVDESS in ACE Lobby ($T_{60}=646$~ms).]{{\includegraphics[width=0.48\columnwidth]{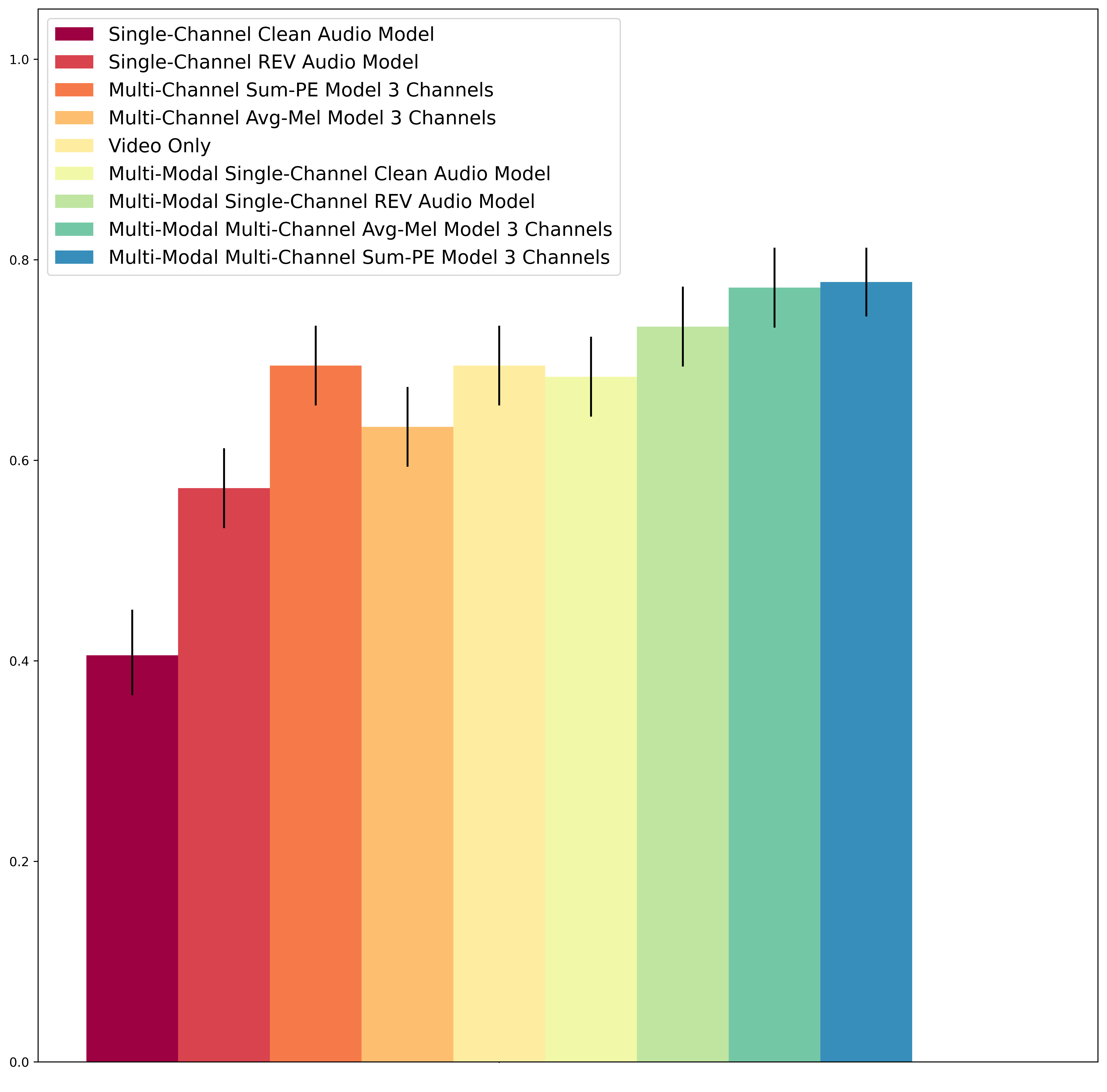} }}%
    \addtolength{\belowcaptionskip}{-3pt}
    \caption{ Accuracy and confidence intervals assessed using the reverberated \ac{RAVDESS} test set for two different rooms from the \ac{ACE} database. 
    }
    \label{Confidence}  
\end{figure}

Analyzing Table~\ref{ACE_results}, it is observed that for the audio-only schemes, the multi-channel processing methods (Avg mel and Sum PE) consistently outperform the single-channel approaches. This is in line with the findings of \cite{ohad2024}. Notably, the multi-modal approaches significantly outperform their uni-modal counterparts.
 These results are also visually demonstrated in Fig.~\ref{Confidence}, demonstrating the advantages of multi-modal processing. 
 \vspace{-.3cm}
 \begin{figure}[h]
\centering
  \includegraphics[width=0.36\textwidth]{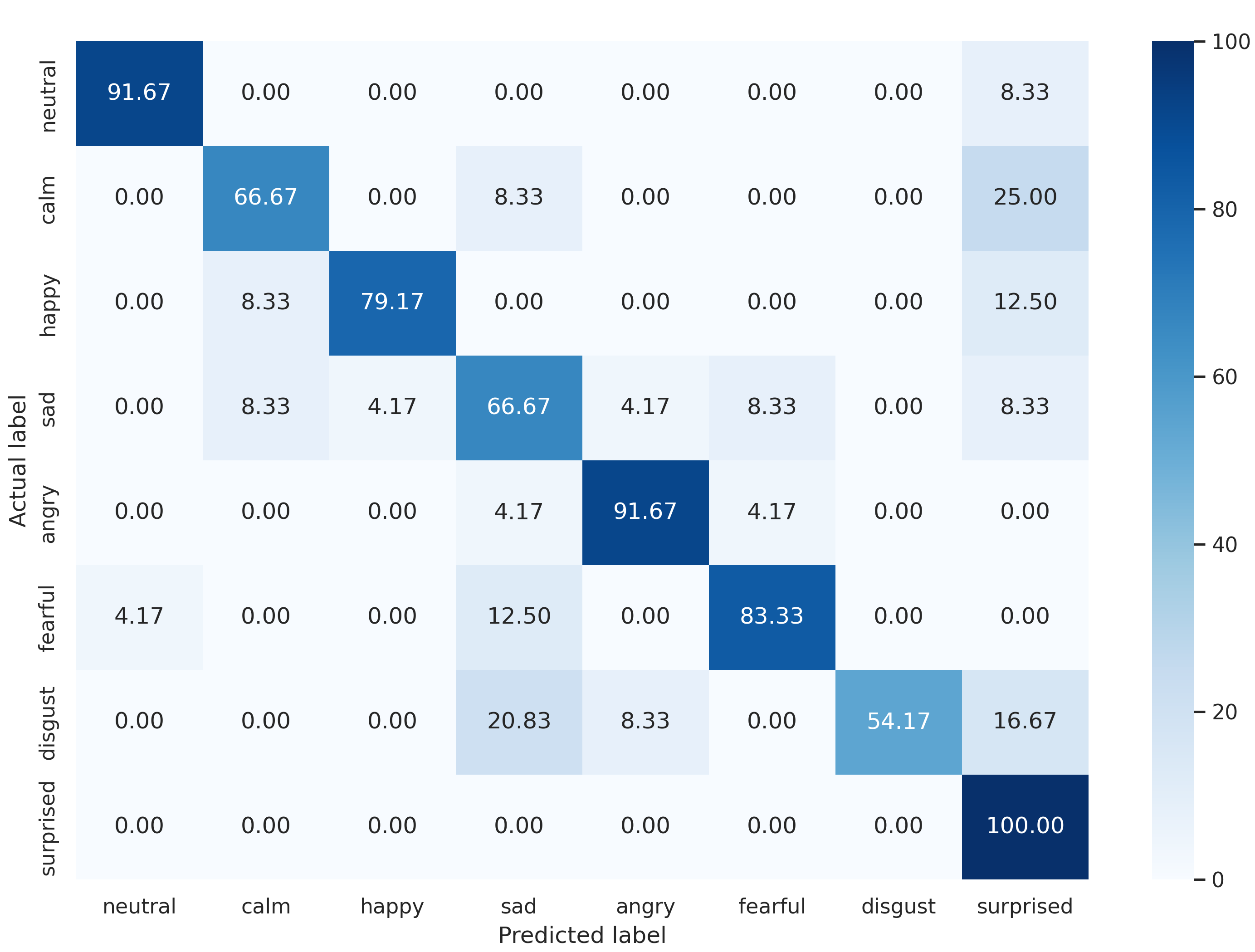}
  \caption{Confusion matrix of the results of multi-channel Sum PE MER model on RAVDESS test set convolved with ACE Lecture Room 2 ($T_{60}=1220$~ms).}
  \label{CM}
\end{figure}
 In addition, Fig.~\ref{CM} presents the confusion matrix for the multi-channel Sum PE \ac{MER} model. The confusion matrix compares the actual target and predicted labels, showing the percentage of correct and incorrect predictions for each class. Beyond measuring accuracy, it also reveals the distribution of errors across different emotions, helping to identify specific misclassifications.

\section{Conclusions}
\label{Conclusions}
In this paper, we presented a \ac{MER} system designed to operate in reverberant and noisy acoustic environments. Our approach demonstrates robust performance across a range of realistic acoustic conditions by combining an extended multi-channel \ac{HTS-AT} for audio processing with an $R(2+1)$D model for video analysis. The \ac{MER} system combines audio and visual modalities, consistently outperforming uni-modal approaches. Using synthetic \acp{RIR} for training and real-world \acp{RIR} from the ACE database for testing, we comprehensively assess our system's performance in diverse acoustic environments. Moreover, the utilization of multi-channel audio processing, particularly the Patch-Embed summation, proves beneficial in mitigating the effects of reverberation and noise over the single-channel case.
This leads to the potential of our \ac{MER} system for applications in various real-world scenarios where acoustic conditions may be far from ideal. Future work could further improve the system's performance in extremely reverberant environments and explore its effectiveness in other emotional datasets.

\balance
\bibliographystyle{IEEEtran}
\bibliography{template}

\end{document}